\pdfoutput=1
\documentclass[aps,pre, twocolumn, groupedaddress]{revtex4-1}

\usepackage{lipsum}
\usepackage{mathtools}
\usepackage{graphicx}
\usepackage{dcolumn}
\usepackage{amsmath}    
\usepackage{amssymb}
\usepackage{bm}
\usepackage{hyperref}
\usepackage{latexsym}
\usepackage{verbatim}
\usepackage[normalem]{ulem}
\usepackage{color}
\usepackage[caption=false]{subfig}
\def\beq{\begin{equation}}
\def\eeq{\end{equation}}

\def\beq{\begin{equation}}                          
\def\eeq{\end{equation}}                          
\def\bea{\begin{eqnarray}}                          
\def\eea{\end{eqnarray}}

\DeclareRobustCommand{\uvec}[1]{{%
  \ifcsname uvec#1\endcsname
     \csname uvec#1\endcsname
   \else
    \bm{\hat{\mathbf{#1}}}%
   \fi
}}
\preprint{}
\begin{document}
\preprint{}
\title{Activity Driven Phase Separation and Ordering Kinetics Of Passive Particles}
\author{Shambhavi Dikshit$^{1}$}
\email{shambhavidikshit.rs.phy18@itbhu.ac.in}
\author{Shradha Mishra$^{1}$}
\email{smishra.phy@iitbhu.ac.in}
\affiliation{$^{1}$Department of Physics, Indian Institute of Technology (BHU), Varanasi, India 221005}
\date{\today}
\begin{abstract}
     The steady state and phase ordering kinetics in a pure active Borwnian particle system 
	are studied in recent years. In binary mixture of active and passive Brownian particles passive particles are used as probe to 
	understand the properties of active medium. 
	In our present study we study the mixture of passive and active Brownian particles. Here we aim to understand the steady state and kinetics of small passive particles in the mixture. In our system,the passive particles are small in size and large in number, whereas ABPs are large in size and small in number. 
	The system is studied  on a two-dimensional substrate using  overdamp Langevin dynamic simulation. The  steady state and kinetics of passive particles are studied for various size 
	and activity of active particles. Passive particles are purely athermal in nature and have dynamics only due
	to bigger ABPs.
	For small size ratio and activity the passive particles remain homogeneous in the system, whereas on increasing
	size ratio and activity they form periodic hexagonal close pack (HCP) spanning clusters in the system. 
	We have also studied the kinetics of growing passive particle clusters. 
	The mass of the largest cluster shows a much  slower growth kinetics in contrast to conserved growth kinetics 
	in ABP system. 
	Our
	study provides an understanding of steady state and kinetics of passive particles in the presence of bigger active particles. The mixture can be thought of as  effect of big  microorganism 
	moving in passive medium.

\end{abstract}
\maketitle
\section{Introduction \label{Introduction}}
Collection of active Brownian particles (ABPs) undergoes a motility induced phase separation without any  cohesion at 
packing density much lower than the  density for phase separation in corresponding passive system \citep{rev, Marchetti, romanczuk, Bechinger, pritha2018}. 
The mechanism of phase
separation is due to the enhanced persistent motion of active particles \citep{cates2015, Alaimo}.\\

 Recent researches have focussed  the kinetics and steady state of pure ABPs or in the mixture 
 of passive particles \citep{Kummel2015, Angelani2011, Leonardo2010, Ray2014, Jay2019}. 
 In a recent work of \cite{Stenhammar2015}, a monodisperse mixture of active passive particles 
 is studied for varying activity and packing fraction of ABPs. Whereas in other studies a field theoretic 
 approach is used to understand the propagation of active passive interface in the mixture of passive and 
 active particles \cite{Wysocki2016}. In the  study of \cite{Wittkowski2017}, a mixture of active passive particle 
 is studied, and different phases and dynamics of system is studied. 
A variety of interesting properties and  phases have been found when ABPs are placed in the mixture with 
passive particles. Asymmetric  passive particles lead to directional
transport and trapping when placed in the sea of ABPs \cite{Pattanayak2019, Malgaretti, Reichhardt, Buttinoni2013}. 
Such systems can be useful  in  industrial and pharmaceutical applications.
Symmetric passive particles in the mixture of ABPs has also been used as a probe to characterise the properties of ABPs \cite{Beatrici2017, Wu1999, Patteson2016}. 
Experiments on the dynamics 
of large passive beads in active bacterial fluid show the persistence motion of bead in bacterial solution \cite{Wu1999}. In the binary mixture of ABPs with passive bead,
large passive particles experience an effective attractive interaction analogous to depletion induced attraction in asymmetric equilibrium binary mixture \cite{Liu2020, pritha2018, Patteson2016}.\\
Most of the above studies of binary active-passive mixture are studied where passive particles are bigger in size and treated 
as a probe to characterise the properties of
active medium \cite{Ni2013, Ni2014, Soft2015} or monodisperse mixture of active passive particles and dynamics of different phases are studied \cite{Stenhammar2015,Wysocki2016,Wittkowski2017}. 
Here we ask the question, how the ABPs which are bigger in size can  influence the characteristics of athermal passive
particles ?
The system resembles  big microorganisms moving in passive fluid. In general thermal and hydrodynamic effects are 
important in normal passive fluid, but we ignore it here to make the model simple and only study the effect of 
activity  on the properties of passive particles. \\ 
We consider a minimal model of mixture of small passive and big active Brownian particles on a two-dimensional substrate. The both types
of particles interact through a short range soft repulsive interaction. The dynamics of active particles
is driven by the active self-propulsion force and interaction with the particles in its surroundings whereas passive 
particles can move only due to the interaction with other particles. The packing fraction of both particles are  same and total packing fraction is kept fixed 
at $0.6$. The dynamics and steady state of passive particles are studied for various size ratios of active  and passive particles and dimensionless activity 
of ABPs. \\
Our main results are as follows: Starting from the random homogeneous mixture of active and passive particles, 
 passive particles start to phase separate with time. The phase separation order parameter of passive particles grows with time and saturates to value $\sim1$ for large 
size ratio and activity and remains much lower than $1$ for small size ratio and activity. Hence a phase diagram is found in the plane of size ratio and activity.
For moderate size ratio and activity the clustered passive particles form hexagonal close packed structures and start to overlap for large size ratio and activity.
The cluster size distribution changes from exponential to power law and power converges to $-2$ for large size ratio and activity. Hence passive particles form the spanning clusters \cite{Rapaport1992} for
larger size ratio and activity. \\
We have further studied the kinetics of the phase separation.  The mass  of the largest cluster grows 
as as a power law with time, with a  much
slower growth kinetics in contrast to equilibirum conserved passive system \cite{modelb}\\

Rest of the article is divided in the following manner. In next section \ref{model}, we describe the model in detail. Section \ref{results} discusses the results of our study and finally 
in section \ref{conclusion} we conclude our results.

\section{Model \label{model}}
\begin{figure} [ht]
\centering
  \includegraphics[scale=0.20]{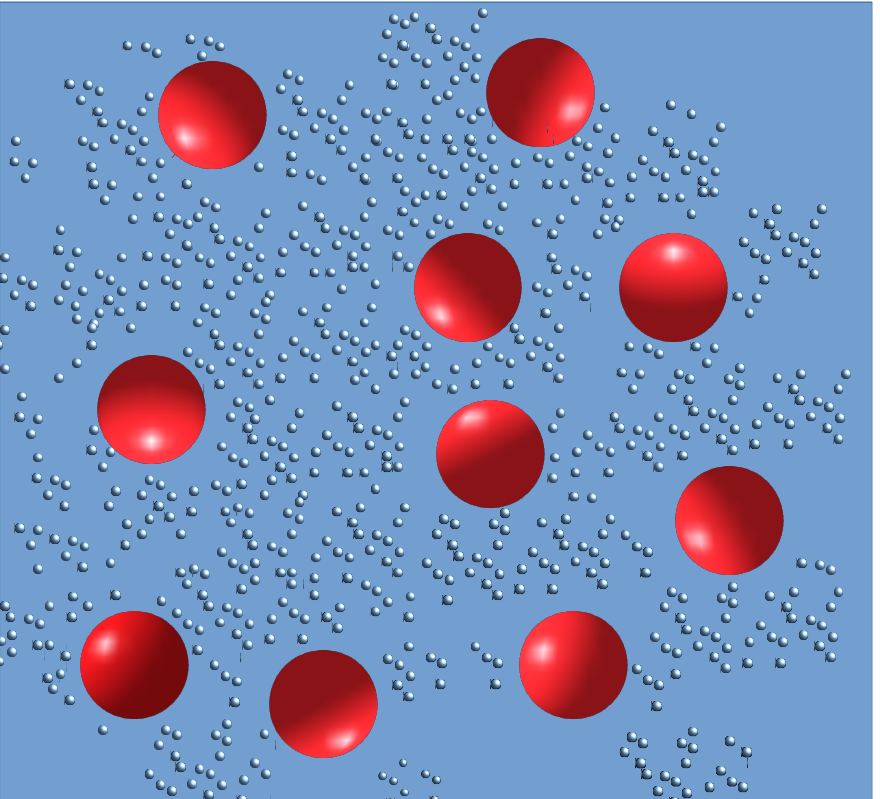}
 
	\caption{(color online) Schematic diagram of the model.	It shows the initial  homogeneous state of the system. 
	Here the  colours and sizes represent two types of particles. Bigger particles (red) are active particles and smaller particles (grey) are passive particles. Small white dot on red particles denote the instantaneous orientation direction $\theta$ of ABPs.}
\label{fig:fig1}
\end{figure}

We consider a binary mixture of  small athermal passive particles in the presence of 
large active Brownian particles (ABPs) on a two-dimensional substrate. 
The active and passive particles are modeled as discs of  radius $r_a$ and $r_p$ respectively. 
We choose $r_a > r_p$,  active particles are larger in size compare to passive particles. 
The  size ratio $S = \frac{r_a}{r_p}$ is one of the control parameters in the model. The radius of the passive particles is kept fixed and radius 
of active particles is tuned to vary the size ratio. We keep the packing fractions of both types of  particles $\phi_a = \frac{\pi r_a^2 N_a}{L^2}$= 
$\phi_p = \frac{\pi r_p^2 N_p}{L^2}$ = $0.3$, hence the number of active particle $N_a$ are less in comparison to that of passive particles $N_p$.
Both active and passive particles are  defined by their position ${\bf r}_i(t)$ and active particles are also having  their orientation $\theta_i(t)$, 
which determines their direction of self-propulsion. They self-propel along their direction of orientation $\theta_i(t)$ with a constant self-propulsion speed
$v_0$. The dynamical Langevin's equations of motion for position and orientation  of active particles are

  \begin{align}
	  \frac{d {\bf r}_{i}^{a}(t)}{d t}=v_{0}\hat{{\bf n}_i}(t)+\mu {{\bf F}}_i(t)
   \label{eq:1}
  \end{align}
 where ${\bf n}_i(t) = (\cos(\theta_i(t)), \sin(\theta_i(t))$ is the unit direction of self-propulsion of ABP. 
 The change in the orientation of the active particle is given by$:$ 
 \begin{align}
  \frac{d\theta_{i}}{dt}=\sqrt{\nu_r}\eta_{i}(t)
  \label{eq:2}
 \end{align}
  here $\eta_i(t)$ is the random Gaussian white noise with mean zero and variance, 
   $<\eta_{i}^{r}(t)\eta_{j}^{r}(t^{'})>=\delta_{ij}\delta(t-t^{'})$, where $\nu_{r}$ 
   is the rotational diffusion constant of active particles.   
  The equation of motion for the passive particle is given as$:$
  \begin{align}
	  \frac{d {\bf r}_{i}^{p}(t)}{d t} =\mu {\bf F}_i(t)
   \label{eq:3}
  \end{align}
  Here, the mobility, $\mu$ is chosen to be the same for  both types of  particles. 
   ${\bf F}_i(t)$ is the force acting on the $i^{th}$ particles, due to all other particles interacting with it
  \begin{equation}
	  {\bf F}_i(t)= \sum_{j\neq i} {\bf F}_{ij}(t)
  \end{equation}
  The force is obtained from the soft-repulsive pair potential ${\bf F}_{ij}= -\nabla U(r_{ij})$, where
	  $U(r_{ij})= K(r_{ij}-(r_{\alpha i}+r_{\alpha' j}))^{2}$ if $r_{ij}\leq(r_{\alpha i}+r_{\alpha' j})$ 
	  and  $0$ otherwise.  $r_{ij}$=$|r_{i}$ - $r_{j}|$
  and $K$ is the force constant and $r_{\alpha}$, is the radius of active or passive particles 
	 for  $\alpha$ and $\alpha'$= $a$ or $p$ respectively.
  $\nu_r^{-1}$ is the time scale  over which the orientation of an active particle changes. Hence,
   $l_p = v_0\nu_r^{-1}$, is the persistence length or run length, is the typical distance
   travelled by an active particle before it changes its direction. In our study, $l_p = (100 r_p$ to $600 r_p)$ is tuned by tuning 
   SPPs $v_0$.
   The $(\mu K)^{-1}= 0.7$ defines the elastic time scale in the system. We define the dimensionless 
   activity $\bar{v}= \frac{l_p}{r_p}$ as the ratio of persistent length to the size of passive particles. The size ratio $S$
   and $\bar{v}$ are the two  tuning parameters in the model. A schematic cartoon of system is shown in Fig. \ref{fig:fig1}, where
   red big particles are ABPs and small gray particles are passive. The white dots on red particles represent their
   instantaneous direction of orientation $\theta$. \\
   We start with random non-overlapping arrangement of active and passive particles 
   on a two dimensional square substrate of linear dimension $L=250 r_p$ with periodic boundary condition.  
   The  equations $\ref{eq:1}-\ref{eq:3}$, are updated and  one simulation step is counted after 
   updation of all the particles once. The time step  $\tau=5 \times10^{-4}$ and 
   total $6\times10^{4}$ simulation steps  are  used to get the  results.  
    We first characterise the effect of size ratio and activity 
   on the steady state of the  athermal passive particles in the mixture. 
	 Then we study the kinetics to the steady state.

\section{Results \label{results}}

\begin{figure*}[ht]
     \centering

       \includegraphics[scale=0.20]{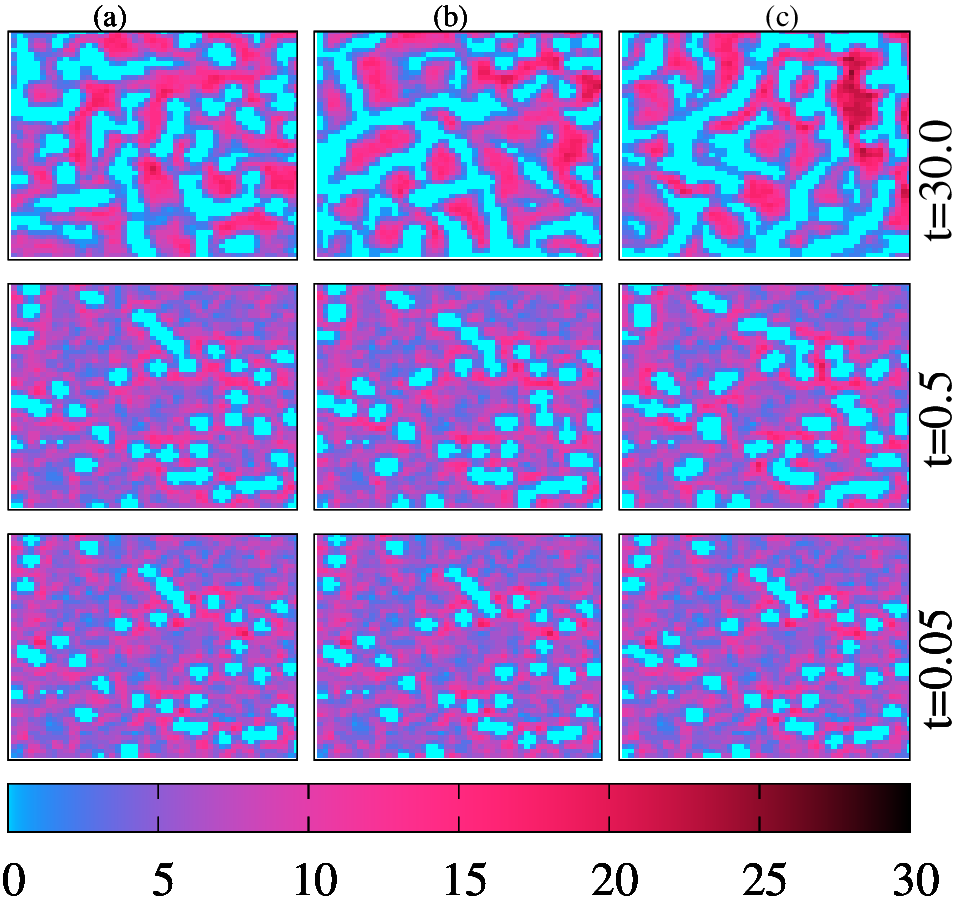}
         \includegraphics[scale=0.20]{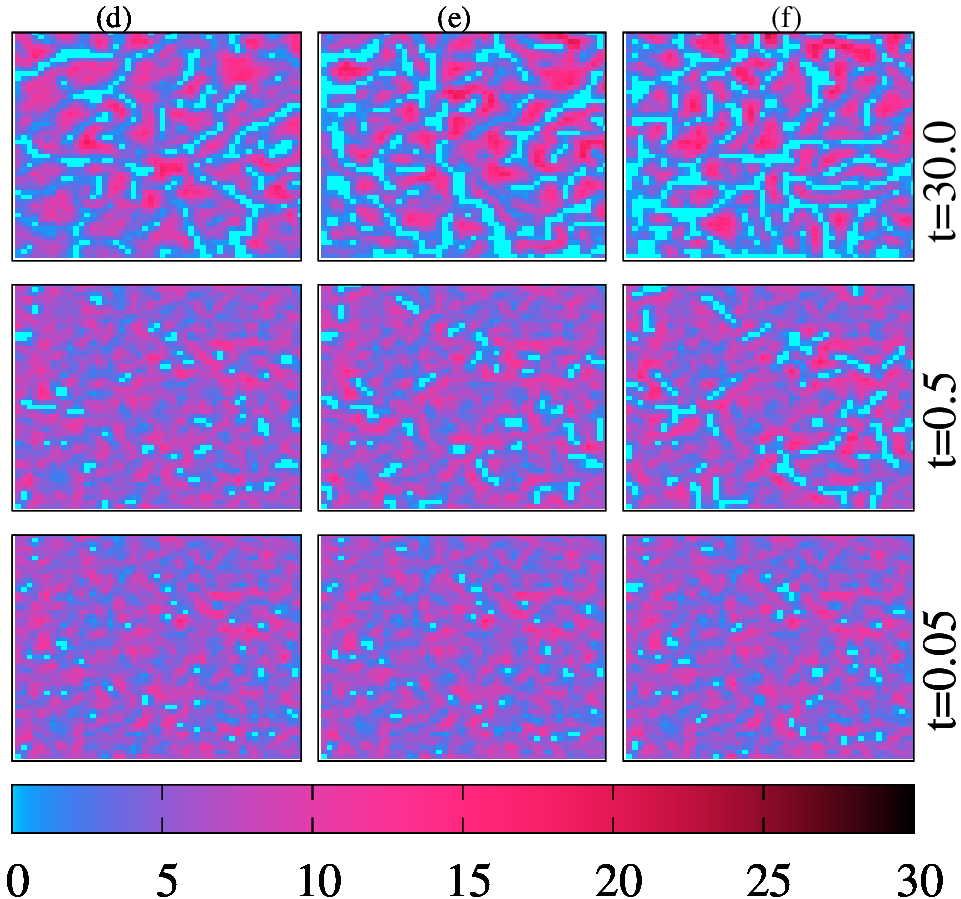}

       \includegraphics[scale=0.175]{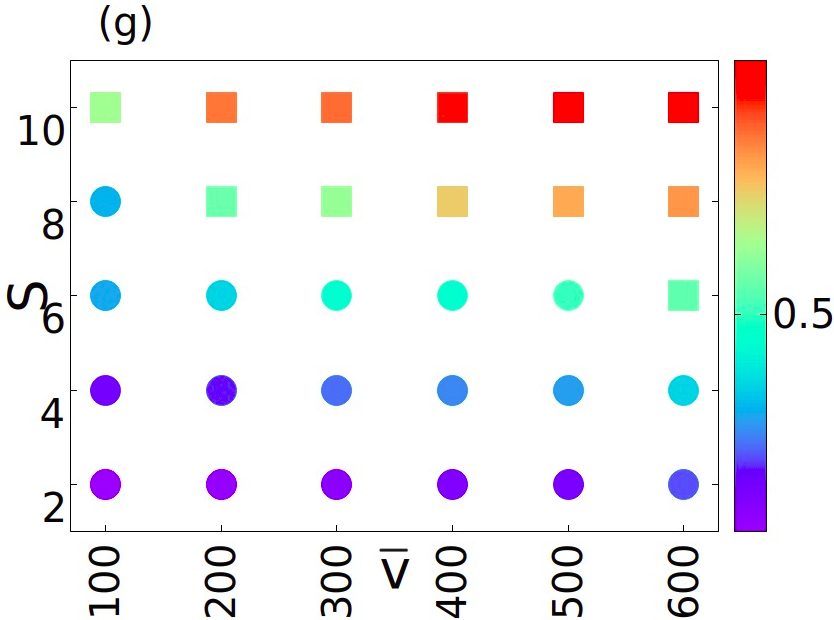}
          \includegraphics[scale=0.205]{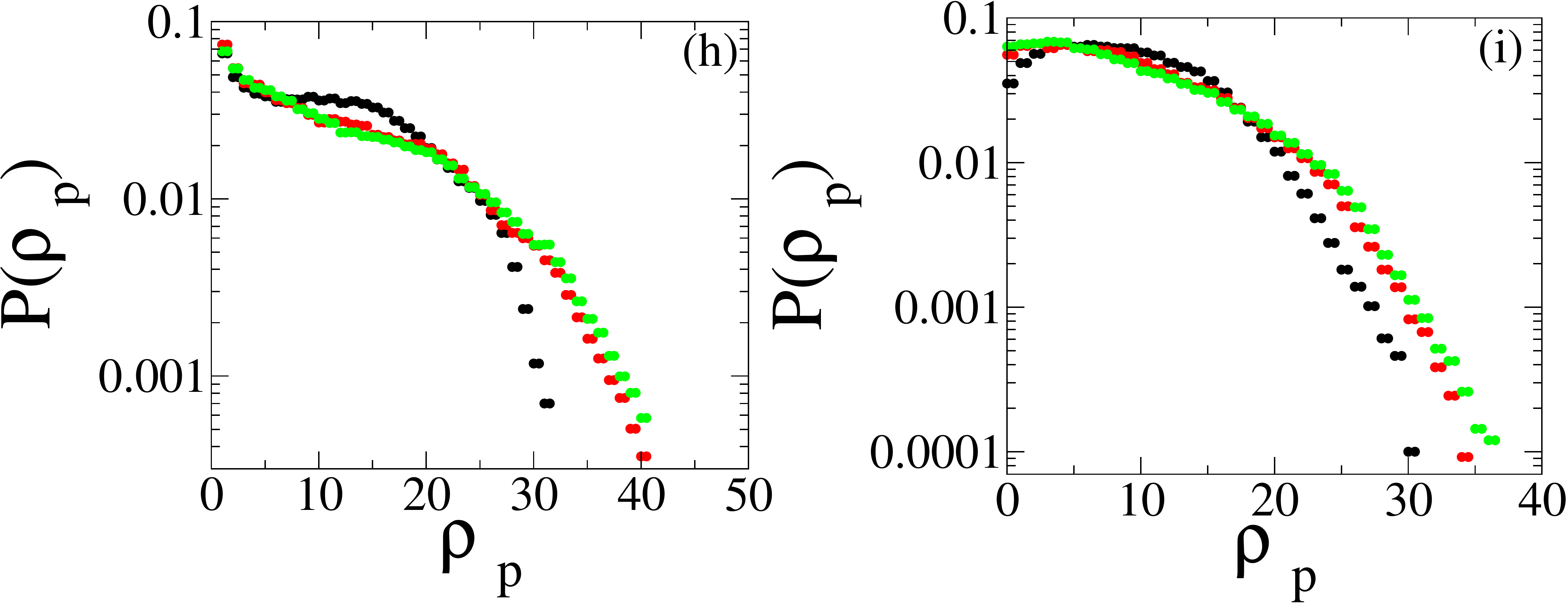}
        
	\caption{ (color online) The real space local density of passive particles at difefrent times. (a-c) are for size ratio $10$ and $\bar{v}$= $100, 300$ and $600$ respectively. (d-f) are for size ratio $6$ and $\bar{v}$= $100, 300$ and $600$ respectively. Lowest panel is for time, $t=0.05$, mid panel is for $t=0.5$ and upper most panel is for $t=30.0$. The lower color plot represents phase separation order parameter {\em (PSOP)} in the plane of size ratio and activity $(S, \bar{v})$.
	Probability distribution function (PDF) of local density of passive particles is plotted for two 
	different size ratios $10$ and $6$ ((h) and (i)) respectively and for three $\bar{v}$= $100, 300$ and $600$.}
        \label{fig:fig2}
\end{figure*}

We start with the random homogeneous distribution of active and passive particles and the equations $\ref{eq:1}-\ref{eq:3}$ are
integrated to update the position and orientation of active and position of passive particles. 
In Fig. \ref{fig:fig2} we plot the time evaluation snapshot
of local density of passive particles at different times $= 0.05$, $0.5$ and $30.0$ and for two different size ratios $S=10$  and $6$  and for $\bar{v}=$ $100$, $300$ and $600$. 
The bright and dark regions show
the lower and higher local density of passive particles  respectively. Local density $\rho_p$ is defined as:  
	 {\em the number of passive particles in the small coarse-grained area ($a=5 r_p \times 5 r_p$)}.  
	 In Fig. \ref{fig:fig2}(h-i) we plot the probability distribution function (PDF) of local density 
	 $P(\rho_p)$ for the same parameters as in Fig. \ref{fig:fig2}(a-f). 
The tail of the distribution is larger for large size ratio $S=10$ and activity $\bar{v}=600$.
As time progresses the passive particle starts to come close to each other. Hence local density $\rho_p$ of passive particle
rich region grows or passive particles phase separate. To characterise the phase separation, 
we calculate the phase separation order parameter {\em (PSOP)}, $\phi(t) = <{(n_{p}(t)-n_{a}(t))}/{(n_{p}(t)+n_{a}(t))}>$, where $n_{p}(t)$ and $n_{a}(t)$ are 
 calculated  as the number of active and passive neighboring particles around a passive particle. $<..>$ means average over all passive particles and $20$
 independent realisations. 
     With time $\phi(t)$  grows  and approaches a  steady state. 
     We calculate the steady state  $\phi = <\phi(t)>_t$, where $<..>_t$ is the average of $\phi(t)$ over a time interval  in the steady state. 
     In Fig. \ref{fig:fig2}(g) we plot the phase diagram in the plane of activity  and size ratio $(S, \bar{v})$. The color shows the magnitude of {\em PSOP}, $\phi(t)$. 
     For large activity and size ratio, phase separation increases in the system.


\section{Characteristics of  steady state}
\subsection{Radial distribution function {\em (RDF)} $g_{pp}(r)$}
\begin{figure*}[ht]
\centering
  \includegraphics[scale=0.30]{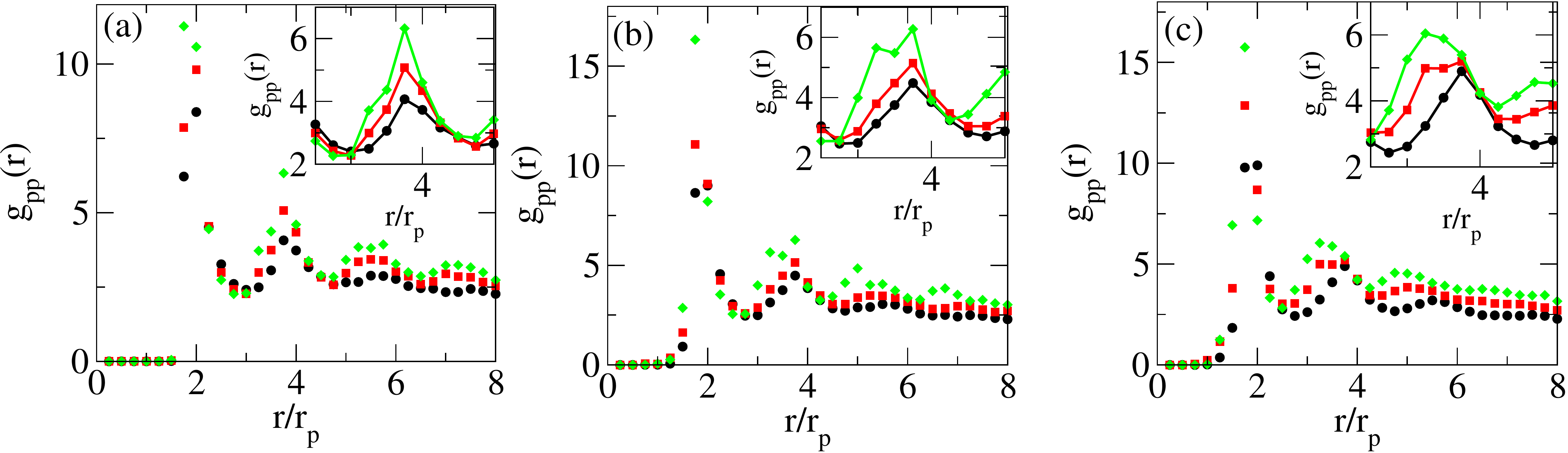}
    \caption{(color online) Passive-passive radial distribution function {\em (RDF)}, $g_{pp}(r)$ is plotted for different parameters. (a-c) is for fixed activity and varying  size ratio. $\bar{v}$ is varied for three different values, $100,300$ and $600$. (a) is for $\bar{v} = 100$ (b) is for $\bar{v} = 300$ and (c) is for $\bar{v} = 600$ and size ratio $S$ is taken $6$, $8$ and $10$ 
 respectively.} 
   \label{fig:fig3}
\end{figure*}

As discussed in previous section, system shows the phase separation for larger size ratio and activity. Hence clustering 
increases with increasing $S$ and $\bar{v}$. We further characterise the structure of the 
clusters  by calculating the radial distribution function {\em (RDF)}
of passive-passive particles for different size ratio $S$ and $\bar{v}$. The {\em RDF}, $g_{pp}(r)$ gives 
the probability of finding a passive particle at a radial distance $r$ from the center of the given passive particle. 
In Figure \ref{fig:fig3} we plot 
the $g_{pp}(r)$ {\em vs.} scaled distance $r/r_p$, for  passive-passive particles for three different activities $\bar{v}=100$, $300$  and $600$ and varying 
the size ratio $S=6$, $8$ and $10$. 
For 
small activity  and $S=6$ as shown in Fig. \ref{fig:fig3}(a), the first peak appears at $r/r_p \approx 2$ and second peak is at $4$ and few more higher order 
peaks are present. But as we increase $S >8$, the location of first peak remains almost the same but a small hump in second peak appears 
at  $r/r_p=2\sqrt{3}$, which is due to the presence of hexagonal close packed structure (HCP) in the clusters. Also for $S>6$, higher 
order peaks are more pronounced. The zoom in plot in Fig. \ref{fig:fig3}(a) (inset) shows the enlarged second peak for three $S=6$, $8$ and $10$. 
As we increase, $\bar{v}=300$, the location of the first peak shifts at distance smaller than $r/r_p \approx 2$, which 
suggests overlapping particles due to soft repulsive interaction. The second peak of $g_{pp}(r)$ appears at
$r/r_p \approx 2\sqrt{3}$, for size ratio $S>6$, hence HCP structure of clusters. Also the distinct higher order
peaks are found for larger size ratio $S>8$. Again the inset Fig.\ref{fig:fig3}(b) shows the zoom in second peak of
the $g_{pp}(r)$. On further increasing $\bar{v}=600$, the first and second peak of $g_{pp}(r)$,
systematically shifts towards the small scaled distance, hence more overlapping particle clusters, the distinct nature of higher
order peaks decreases on increasing $S$.  Hence periodicity of clusters weakens on increasing $S$ for large activity $\bar{v}>300$. The inset Fig. \ref{fig:fig3}(c)
shows the zoom in second peak, where the structure in second peak has been disappeared for large activity $\bar{v}=600$. \par
Hence using {\em RDF}, we find the periodic $HCP$ nature of particle clusters first  increases on increasing activity and size ratio and
again for very large $\bar{v}=600$ and $S=10$, overlapping large clusters are found with weaker HCP structure. Now we further characterise 
the characteristics of large clusters by calculating the cluster size distribution {\em (CSD)} of different size clusters.

\subsection{Cluster size distribution {\em (CSD)}}
\begin{figure*}[ht]
\centering
 \includegraphics[scale=0.30]{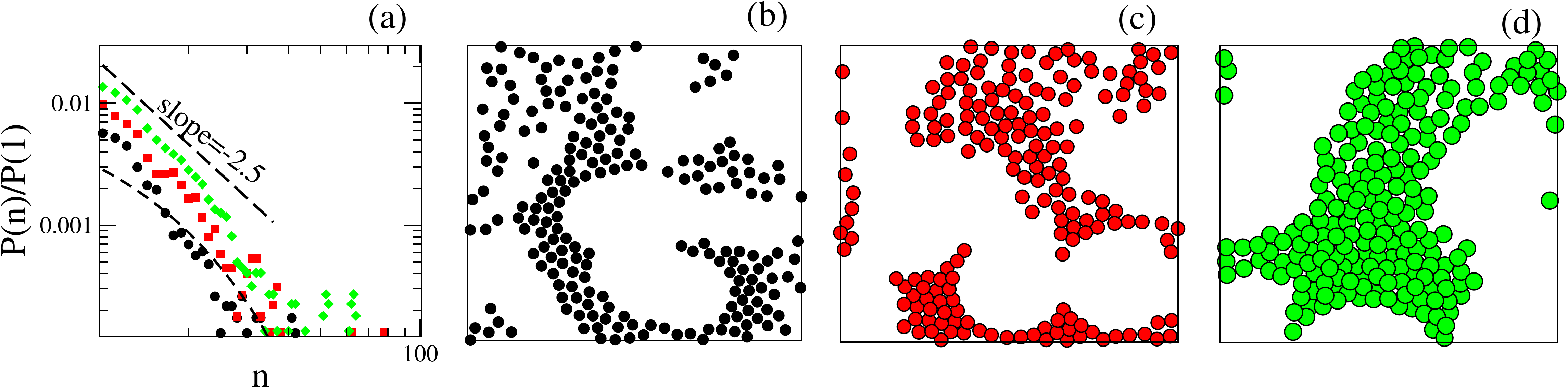}
  \includegraphics[scale=0.30]{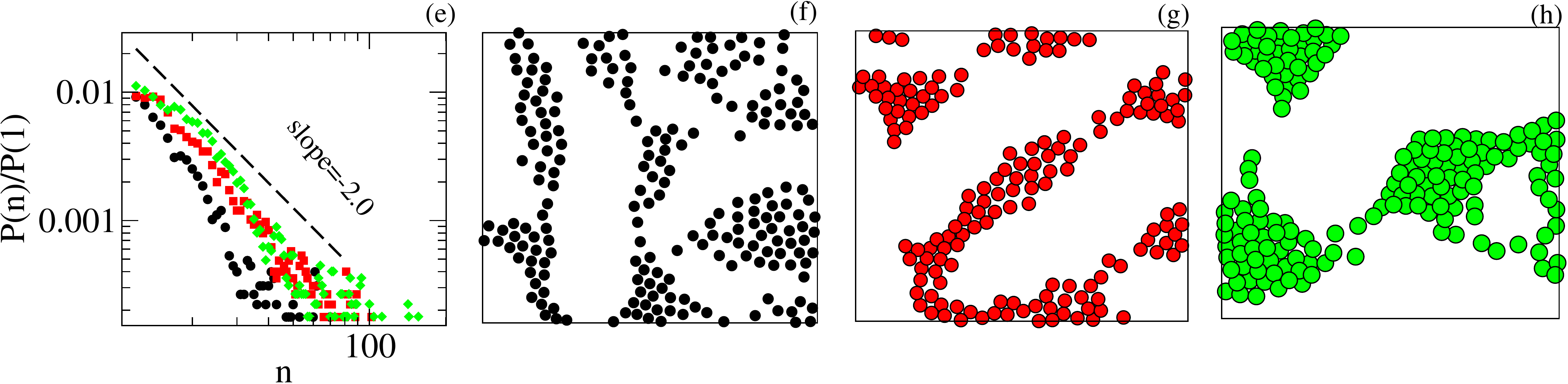}

 \includegraphics[scale=0.30]{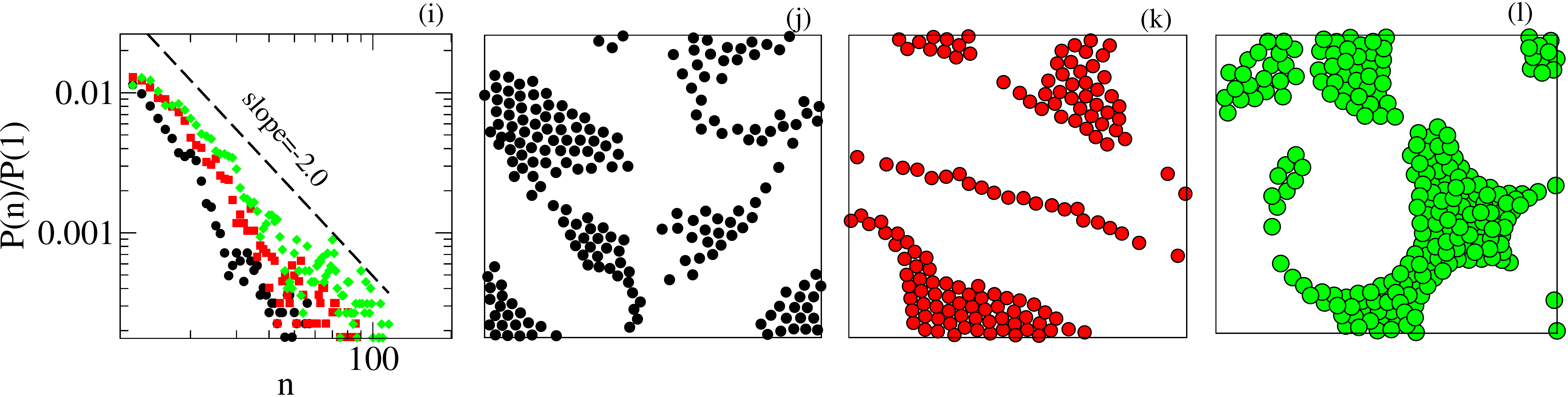}
	\caption{(color online) Cluster size distribution {\em (CSD)} and real space snapshots of part of the system are plotted for different parameters. Different vertical panels (a-d) are for fixed $\bar{v} =100$, (e-h) are for  $\bar{v}= 300$ and (i-l) are for $\bar{v} =600$  and black, red and green colors are for size ratios $6,8$ and $10$ respectively. {\em CSD} is shown in (a),(e) and (i) for $\bar{v}= 100,300$ and $600$ respectively and for size ratio $6,8$ and  $10$. Symbols have the same meaning as in Fig. \ref{fig:fig3}} 
 \label{fig:fig4}
\end{figure*}

To further understand the characteristics of clusters we calculate the probability distribution function of different sized clusters. A cluster is defined as set of particles connected 
by a  distance smaller or equal to $ 2r_p$ {\em (diameter of the particle)}. 
A cluster of size $n$ has $n$ $-${\em particles connected  cluster}. 
Then we calculate the number of different sized 
cluster in the total system. In this manner, all particles are part of a single particle cluster. Hence number of clusters of size $n=1$ is total number of particles.
We further calculate the fraction of cluster of size $n$, or cluster size distribution {\em (CSD)}.
In Fig. \ref{fig:fig4} we plot the normalised  cluster size distribution {\em (CSD)} $P(n)/P(1)$, where $P(i)$ is obtained from the counting all the clusters of size 
$i=1,2,...,n$. 
In Fig. \ref{fig:fig4}(a) we plot the {\em CSD} for activity $\bar{v}=100$ for three different size
ratios  $S=6$, $8$ and $10$. For small activity, $\bar{v}<300$ and size ratio $S <8$, the {\em CSD} decay exponentially at  $n>20$. Hence for small 
activity and size ratio we find small clusters and particles are well separated from each other (as shown in Fig. \ref{fig:fig4}(b)). Hence  weak 
clusters as
 found in the {\em RDF}, $g_{pp}(r)$ plot as shown in Fig. \ref{fig:fig3}(a). As we increase
 size ratio $S>6$,
  for $\bar{v}=100$, clustering increases and {\em CSD} decay as a power law. The power approaches $-2.5$ for large size ratio $S=10$.
The real space snapshots of particles show the enhanced clustering on increasing $S$. For moderate activity 
$\bar{v}=300$, {\em CSD}  for small size ratio $S=6$, is exponential with large exponential tail and for large size ratio
$S>6$, the exponential tail approaches  to power law tail $-2$. Also the real space snapshots show the clustering increases on increasing $S$. 
But for larger size ratio
$S=10$, particles start to overlap and HCP structure weakens as shown in Fig. \ref{fig:fig3}(b). As we further increase activity 
$\bar{v}=600$, the {\em CSD} is power law for all size ratios $S=6$, $8$ and $10$ and power slowly converges to $-2$ for largest size ratio $10$. The power $-2$ suggest that for large activity passive particles form the 
percolating clusters spanning the whole system \cite{Rapaport1992}.\\

\section{Growth kinetics}
\begin{figure}[hbt]
 \centering
  \includegraphics[scale=0.20]{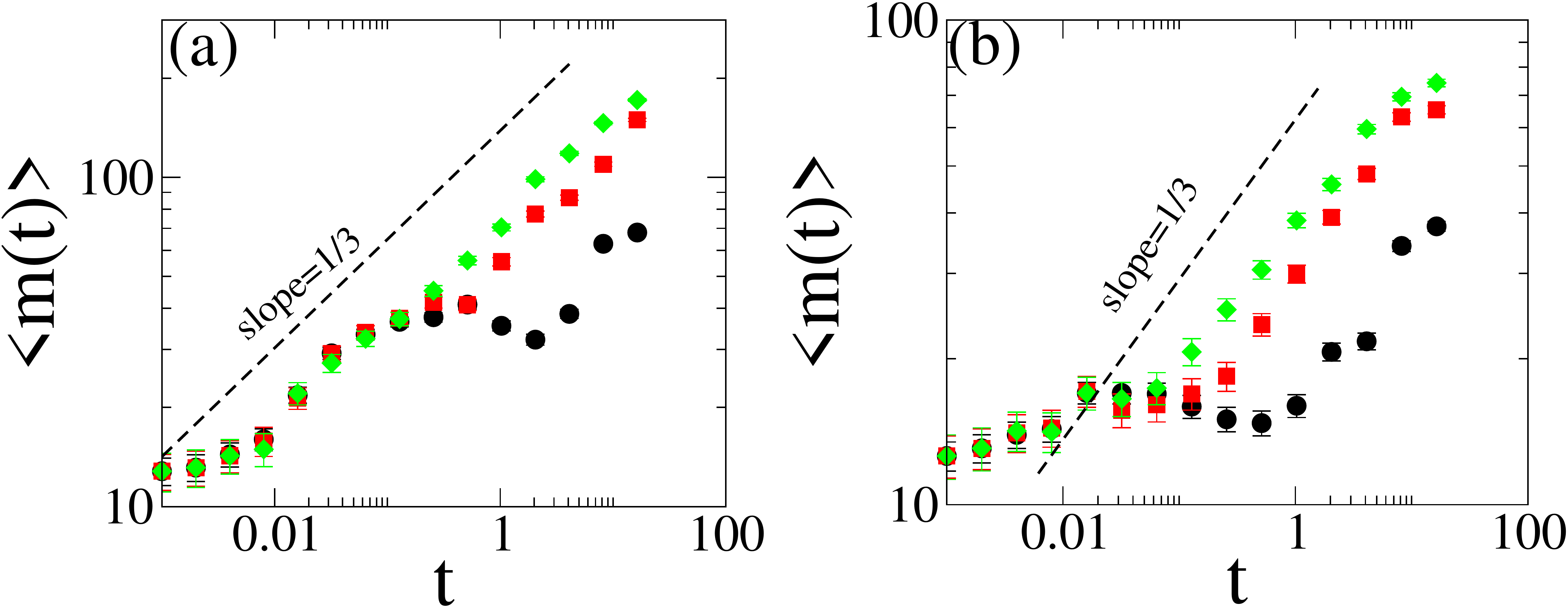}

	\caption{(color online) Mean mass  of the largest cluster $<m(t)>$ is plotted for activities $\bar{v}= 100,300$ and $600$. (a) is for size ratio $10$ and (b) is for size ratio $6$. Black, red and green show the activities $100, 300$ and $600$ respectively. Dashed line is the line with slope $1/3$}
          \label{fig:fig5}
 \end{figure}

After understanding the steady state properties of the mixture, we study the kinetics of phase separation of athermal passive particles.
We  study the  kinetics of growing cluster by calculating the mass of the largest  cluster at different times, $m(t)$.
In Figure \ref{fig:fig5}, we plot the mean  mass of the largest cluster $<m(t)>$ for two different size ratios $S=10$ and $6$ and for different $\bar{v}$. 
where $<..>$ is mean over $50$ independent realisations.
In Fig. \ref{fig:fig5}(a) we plot $<m(t)>$ for size ratio $10$ for three different activities $\bar{v}=100, 300$ and $600$. For all activities and $S=10$, at very early time $t<10^{-2}$,
$m(t)$ grows with time and then for an intermediate times $t~ (10^{-2}, 10)$, growth becomes slow and $<m(t)>$ develops  a plateau and again 
starts to grow for late times $t \ge 10$. The plateaus region decreases 
with increasing activity. The late time growth of $<m(t)> \sim t^{1/3}$ for all activities. In Fig. \ref{fig:fig5}(b) we plot
the $<m(t)>$ vs. $t$ for  size ratio $S=6$. The growth of $<m(t)>$ shows the same behaviour as for $S=10$, only the  plateaus region is increased  for $S=6$. For 
largest activity $\bar{v}$ and size ratio $S=10$, $<m(t)>$ increases monotonically with time $t$ with $m(t) \sim t^{1/3}$. \par


\section{Conclusion \label{conclusion}}
We have studied the phase separation and ordering kinetics of a binary mixture of passive and active Brownian particles on a two-dimensional substrate with 
periodic boundary condition. The passive particles are smaller in size comparison to the ABPs. The ABP moves along the direction of their heading and both 
types of particles interact through a short range soft core repulsive interaction. Hence the dynamics of passive particles are only due to interaction force
among the particles. The system is studied for various size ratios $S$ and activities $\bar{v}$ of active particles. We focus our study on the steady state
and kinetics of small passive particles in the presence of big passive particles.
Starting from the random homogeneous state,  the clustering of small passive particle is measured by calculating phase separation order parameter {\em (PSOP)}. 
The {\em PSOP} is small for size ratio $S<6$ and small activity $\bar{v}<300$, whereas for large size ratio $S>8$ and higher activity $\bar{v}>300$,
{\em PSOP} approaches $\sim1$. The clusters of passive particles are random small clusters for small size ratio and activity and HCP structures are formed for 
intermediate size ratio $S=8$ and activity $\bar{v}=300$ and then overlapping clusters are found for large size ratio and activity $S>8$ and $\bar{v}=300$.
The Cluster size distribution  decays exponentially for small size ratio and activity and approaches to a power law decay with exponent $-2$ for large 
size ratio and activity. The power law decay with power $-2$ indicates the formation of connected clusters as in \cite{Rapaport1992} for large size ratios and activities. \\
We have also calculated the kinetics of growing cluster of passive particles. The mean mass of the largest cluster grows with time as a power law. {\bf The  growth law is much smaller than the corresponding equilibirum
system with conserved growth kinetics} \cite{cates2013,wittkowski2014,ajbray,Pattanayak,Pattanayak2020}.               \\
Hence our study gives the steady state of collection of passive particles moving under the effect of dynamics of active Brownian particle.It focuses on the steady state and kinetics for binary mixture where passive particles are much smaller than the ABPs. The system resembles the effect of big microorganism moving in 
passive medium.

\section{conflicts of interest \label{conflicts of interest}}
There are no conflicts of interest to  declare.

\section{Acknowledgement \label{Acknowledgement}}

The
support and the resources provided by PARAM Shivay
Facility under the National Supercomputing Mission,
Government of India at the Indian Institute of Technology, Varanasi are gratefully acknowledged. Computing facility at Indian Institute of Technology(BHU),
Varanasi is gratefully acknowledged.


\begin{thebibliography}{9}
\bibitem{rev} S. Ramaswamy,  Annu. Rev. Condens. Matter Phys.,{\bf 1},323(2010).

\bibitem{Marchetti} M. C. Marchetti, J. F. Joanny, S. Ramaswamy, T. B.
Liverpool, J. Prost, Madan Rao, R. Aditi Simha, Rev. Mod. Phys.,{\bf 85}, 1143(2013).

\bibitem{romanczuk}  P. Romanczuk, M. Bär, W. Ebeling, B. Lindner, L.
Schimansky-Geier, Eur. Phys. J. ST,{\bf 1} 202(2012).

\bibitem{Bechinger} C. Bechinger, R. D. Leonardo,H. Lowen, C. Reichhardt, G. Volpe,G. Volpe, Rev. Mod. Phys.,{\bf 88}, 045006(2016)

\bibitem{pritha2018} P.Dolai, A.Simha, S.Mishra, Soft Matter(2018)

\bibitem{cates2015} M. E. Cates and J. Tailleur,Annu. Rev. Condens. Matter Phys. (2015)

\bibitem{Alaimo}  F. Alaimo and A. Voigt, Phys. Rev. E. {\bf 98} ,  032605 (2018) 











\bibitem{Kummel2015} F. Kummel,P. Shabestari,C. Lozano, G. Volpe and
C. Bechinger, Soft Matter (2015)

\bibitem{Angelani2011} L. Angelani, C. Maggi, M. L. Bernardini, A. Rizzo, and R. Di Leonardo,  Phys. Rev.Lett. {\bf 107}, 138302 (2011)

\bibitem{Leonardo2010} R. Di Leonardo, L. Angelani, D. Dell’Arciprete , G. Ruocco, V. Iebba , S. Schippa , M. P. Conte , F. Mecarini ,
F. De Angelis and E. Di Fabrizio, PNAS {\bf vol. 107}, {\bf no. 21} , 9541–9545 , (2010)

\bibitem{Ray2014} D. Ray, C. Reichhardt and C. J. Olson Reichhardt, Phys. Rev.E. {\bf 70}, 013019, (2014)

\bibitem{Jay2019} J. Prakash  Singh and  Shradha  Mishra†, Physica A., vol. {\bf 544} (2020)

\bibitem{Stenhammar2015} J. Stenhammar, R. Wittkowski, D. Marenduzzo and M. E. Cates, Phys. Rev. Lett. {\bf 114} 018301 (2015)

\bibitem{Wysocki2016} A. Wysocki, R. G Winkler and G. Gompper, New J. Phys. {\bf 18} 123030 (2016)

\bibitem{Wittkowski2017} R. Wittkowski, J.Stenhammar and M. E. Cates,  New J. Phys. {\bf 19} 105003 (2017)



\bibitem{Pattanayak2019}  S. Pattanayak, R. Das, M. Kumar and S. Mishra, Eur. Phys. J. E {\bf 42},{\bf 62} (2019)

\bibitem{Malgaretti}  P. Malgaretti and H. Stark, J. Chem. Phys.146, 174901 (2017)

\bibitem{Reichhardt}  C.  Reichhardt  and  C.J.  Olson  Reichhardt,  Phys.  Rev.  E{\bf 97},  052613(2018)

\bibitem{Buttinoni2013} I. Buttinoni,J. Bialké , F. Kümmel,H. Löwen, C. Bechinger and T. Speck, Phys. Rev. Lett. {\bf 110} , 238301 (2013)

\bibitem{Beatrici2017} C. P. Beatrici,R. M. C. de Almeida, L. G. Brunnet, Phys. Rev. E. {\bf 95} 032402 (2017)

\bibitem{Wu1999}    Xiao-L. Wu  and A. Libchaber,  Phys. Rev. Lett. {\bf vol. 84}, (1999)

\bibitem{Patteson2016}  A. E. Patteson, A. Gopinath, P. K. Purohit and  P. E. Arratia, Soft Matter, {\bf 12}, 2365 (2016)

\bibitem{Liu2020}   P. Liu, S. Ye, F. Ye, K. Chen, and M. Yang, Phys. Rev. Lett. {\bf 124}, 158001 (2020)


\bibitem{Ni2013} R. Ni, M. A. Cohen Stuart and M. Dijkstra, ncomms {\bf 4} , 2704(2013)

\bibitem{Ni2014} R. Ni, M. A. C. Stuart, M. Dijkstra  and P. G. Bolhuis, Soft Matter {\bf 10}, 6609-6613 ,(2014)

\bibitem{Soft2015}  F. Kümmel, P. Shabestari, C. Lozano, G. Volpe cd and C. Bechinger , Soft Matter, {\bf 11} ,  6187-6191, 2015

\bibitem{Rapaport1992} D. C. Rapaport, Journal of Statistical Physics, {\bf Vol. 66}, 679-682, (1992)


\bibitem{modelb} S. Puri and V. Wadhawan(eds.), Kinetics of Phase Transition, CRC Press,Boac Raton (2009)

\bibitem{cates2013} J. Stenhammar, A. Tiribocchi, R. J. Allen, D. Marenduzzo, and Michael E. Cates, Phys. Rev.Lett. {\bf 111} , 145702 , (2013)

\bibitem{wittkowski2014} R. Wittkowski, A. Tiribocchi, J. Stenhammar, R. J. Allen,
D. Marenduzzo and M. E. Cates, ncomms {\bf 5351} (2014)

\bibitem{ajbray} Theory of phase-ordering kinetics, A.J. Bray, Advances in Physics, vol. {\bf 43} ,(1994)

\bibitem{Pattanayak} S. Pattanayak, S. Mishra, and S. Puri, Cond. Matt. Soft(2021) 

\bibitem{Pattanayak2020} S. Pattanayak , J. Prakash Singh, M. Kumar and S. Mishra,Phys. Rev. E.{\bf 101},  052602 (2020)






 


































\end{thebibliography}
\end{document}